\documentstyle[12pt]{article}
\date{}
\begin{document}
\title{On the Four--Photon Decay of the Neutral Pion}
\protect\author{E.L.~Bratkovskaya, E.A.~Kuraev, Z.K.~Silagadze\thanks
{ permanent address: Budker Institute of Nuclear Physics,
630 090, Novosibirsk, Russia.}
\vspace*{3mm} \\
\small \em  Bogoliubov Laboratory of Theoretical Physics,
Joint Institute for Nuclear Research,\\
\small \em 141980 Dubna, Moscow Region,  RUSSIA \vspace*{3mm} \\}
\maketitle
\begin{abstract}
The dominant contribution to the $\pi^0 \to 4\gamma$ branching ratio,
coming from purely electromagnetic photon-splitting graph, is calculated.
The result $Br(\pi^0 \to 4\gamma) \simeq (2.6 \pm 0.1) \cdot 10^{-11}$
is about three orders of magnitude below the present experimental limit.
\end{abstract}

Although a C-violating decay $\pi^0\to 3\gamma$ is expected to have an
extremely small branching ratio \cite{b1}, beyond the reach of any
present or future experimental facilities, its experimental study has
attracted considerable attention \cite{b2} because any observed anomaly
in this process would be a clear signal of a new physics.

Any $\pi^0 \to 3\gamma$ searching experiment has as a by-product,
information about the allowed decay $\pi^0\to 4\gamma$, that is a
potential background for $\pi^0\to 3\gamma$. The experimental upper
limit on the branching ratio $Br(\pi^0\to 4\gamma)$ was gradually
improved \cite{b3,b4,b5} and lowered up to $2\cdot 10^{-8}$
in \cite{b2}.  Some theoretical estimates can be found in the
literature for $Br(\pi^0\to 4\gamma)$ \cite{b6,b7,b8}, with rather
broad ranges from $10^{-9}$ to $10^{-16}$. In our opinion, the results
of \cite{b8} are more reliable, the authors giving the most thorough
investigation of the subject.

As argued in \cite{b8}, the dominant contribution to the $\pi_0\to
4\gamma$ branching ratio is expected to come from the purely
electromagnetic photon--splitting graph of Fig.~\ref{fig1},
contributions from any hadronic intermediate states being less
significant, especially for a PCAC--satisfying models.
The calculation of this contribution is still absent to our best
knowledge, and will be performed in the present note.

Using the standard covariant phase-space calculation technique \cite{b9}
and factoring out some numerical constants from the decay amplitude,
we can write
\begin{eqnarray}
Br(\pi^0\to 4\gamma) \simeq {\Gamma(\pi^0\to 4\gamma)\over
\Gamma(\pi^0\to 2\gamma)} = {1\over 6\pi}
\left({\alpha\over 8\pi}\right)^4 \ R,
\label{eq1}\end{eqnarray}
where
\begin{eqnarray}
R = \int\limits_0^1 d s_1 \int\limits_0^{s_1} d s_2
\int\limits_{s_2/s_1}^{1-s_1+s_2} {d u_1\over \sqrt{\lambda(1,s_2,s_2
^\prime)}}
\int\limits_{u_2^-}^{u_2^+} d u_2 \int\limits_{-1}^1 {d \zeta\over
\sqrt{1-\zeta^2}} \ F(s_1,s_2,u_1,u_2,t_2(\zeta)).
\label{eq2}\end{eqnarray}
In (\ref{eq2}) we have introduced a demensionless version of Kumar's
invariant variables \cite{b9}.
\begin{eqnarray}
s_1 = {1\over m^2} (q-k_1)^2, \ \
s_2 = {1\over m^2} (q-k_1-k_2)^2, \ \
u_1 = {1\over m^2} (q-k_2)^2, \ \
u_2 = {1\over m^2} (q-k_3)^2,
\label{eq3}\end{eqnarray}
$m$ being the pion mass.

One more invariant variable $t_2={1\over m^2} (q-k_2-k_3)^2$ is
a linear function of the integration variable $\zeta$
\begin{eqnarray}
t_2=u_1-{1\over 2} (1+u_1)(1-u_2) -{1\over 2} (1-u_1)(1-u_2)
\left[ \xi \eta - \sqrt{(1-\xi^2)(1-\eta^2)} \zeta\right],
\label{eq4}\end{eqnarray}
where
\begin{eqnarray}
\xi = {\lambda(1,s_2,s_2^\prime) - (1-s_1)^2 + (1-u_1)^2 \over
2 (1-u_1) \sqrt{\lambda(1,s_2,s_2^\prime)} }, \ \
\eta = {(1-s_3^\prime)^2 - (1-u_2)^2 - \lambda(1,s_2,s_2^\prime) \over
2 (1-u_2) \sqrt{\lambda(1,s_2,s_2^\prime)} },
\label{eq5}\end{eqnarray}
$\lambda(x,y,z) = x^2+y^2+z^2-2(xy+xz+yz)$ is a conventional triangle
function, and
\begin{eqnarray}
s_2^\prime = 1+s_2 -u_1 -s_1, \ \ s_3^\prime = 2 - s_1 - u_1 - u_2.
\label{eq6}\end{eqnarray}
At last, the limits of integration for the $u_2$-variable in (\ref{eq2})
are
\begin{eqnarray}
u_2^\pm = 1 - {1\over 2}(u_1+s_1) \pm {1\over 2}
\sqrt{\lambda(1,s_2,s_2^\prime)}.
\label{eq6_1}\end{eqnarray}
Function F stands for a half sum of the squared helicity amplitudes
\begin{eqnarray}
F = \sum\limits_{\{\lambda\}} |M_{+\lambda_2\lambda_3\lambda_4}|^2.
\label{eq7}\end{eqnarray}
To evaluate these helicity amplitudes, it is convenient to use the
light-light scattering tensor from \cite{b10}. In fact, as  virtuality
of the intermediate photon is $\sim m$ and much bigger than the
electron mass, we have used the asymptotic form of the light-light
scattering amplitudes for the massless electron in the loop
\cite{b11,b12}.

Let us note, however, that we can not use the polarization vectors from
\cite{b10} because of an additional photon and the need for photon
permutations. Instead we have taken polarization vectors in the form
which appeared useful in various QED calculations \cite{b13}:
\begin{eqnarray}
&&\varepsilon_\mu^{(\lambda_m)} (k) = N_m \left[q^{(m)}\cdot k \ p_\mu^
{(m)}
- p^{(m)}\cdot k \ q_\mu^{(m)} + i \lambda_m \varepsilon_{\mu\nu\lambda
\sigma}
p^{(m)\nu} \ k^\lambda \ q^{(m)\sigma}\right] \nonumber\\
&& N_m^{-1} = 2 \sqrt{p^{(m)}\cdot q^{(m)} \ p^{(m)}\cdot k \ q^{(m)}
\cdot k},
\ \ m=1\div 4, \ \ {p^{(m)}}^2 = {q^{(m)}}^2 = 0.
\label{eq8}\end{eqnarray}
Where for various photons we take
\begin{eqnarray}
p^{(1)} = k_2, \ \ p^{(2)} = k_3, \ \ p^{(3)} = k_4, \ \ p^{(4)} = k_1,
\nonumber \\
q^{(1)} = k_4, \ \ q^{(2)} = k_1, \ \ q^{(3)} = k_2, \ \ q^{(4)} = k_3.
\label{eq9}\end{eqnarray}
The polarization vectors from \cite{b10} can also be expressed in this
form $(8\Delta = k_2\cdot k_3 \ k_3\cdot k_4 \ k_2\cdot k_4)$:
\begin{eqnarray}
&& {u_\mu^\prime }^{(-\lambda_2)} = {1\over 4 \sqrt{2\Delta}}
[k_4\cdot k_2 \ {k_3}_\mu - k_3\cdot k_2 \ {k_4}_\mu + i \lambda_2
\varepsilon_{\mu\nu\lambda\sigma} k_3^\nu k_2^\lambda k_4^\sigma],
\nonumber\\
&& u_\mu^{(\lambda_3)} = {1\over 4 \sqrt{2\Delta}}
[k_2\cdot k_3 \ {k_4}_\mu - k_4\cdot k_3 \ {k_2}_\mu + i \lambda_3
\varepsilon_{\mu\nu\lambda\sigma} k_4^\nu k_3^\lambda k_2^\sigma] \equiv
\varepsilon_\mu^{(\lambda_3)}, \label{eq10} \\
&& u_\mu^{(-\lambda_4)} = {1\over 4 \sqrt{2\Delta}}
[k_3\cdot k_4 \ {k_2}_\mu - k_2\cdot k_4 \ {k_3}_\mu + i \lambda_4
\varepsilon_{\mu\nu\lambda\sigma} k_2^\nu k_4^\lambda k_3^\sigma].
\nonumber\end{eqnarray}
But ${u_\mu^\prime}^{(-\lambda_2)}$ and $u_\mu^{(-\lambda_4)}$ differ by
the phase factors from $\varepsilon_\mu^{(\lambda_2)}$ and
$\varepsilon_\mu^{(\lambda_4)}$ (note that, in contrast with \cite{b10},
${u_\mu^\prime}^{(-\lambda_2)}$ corresponds to the $+\lambda_2$ circular
polarization for the second photon because now it is also outgoing).
Therefore, while using the expressions from \cite{b10}, we should not
forget the relevant phase factors. For example,
\begin{eqnarray}
{u^\prime}^{(\lambda_2)}\cdot \varepsilon^{(\lambda_2)} =
{N_2 \  k_2\cdot k_3\over 2\sqrt{2\Delta}} \Phi (\lambda_2; 1234),
\nonumber\end{eqnarray}
where
\begin{eqnarray}
&& \Phi (\lambda; 1234) = k_1\cdot k_2 \ k_3 \cdot k_4
+ k_1\cdot k_3 \ k_2 \cdot k_4 -  k_1\cdot k_4 \ k_2 \cdot k_3 +
i \lambda \ [k_1,k_2,k_3,k_4], \nonumber\\
&& [k_1,k_2,k_3,k_4] = \varepsilon_{\mu\nu\lambda\sigma} k_1^\mu k_2^\nu
k_3^\lambda k_4^\sigma.
\label{eq11}\end{eqnarray}
Owing to the remarkable cyclic symmetry in the definition (\ref{eq8}),
(\ref{eq9}) of the $\varepsilon_\mu^{(\lambda_i)}$ polarization vectors,
for the $\pi^0 \to 4\gamma$ helicity amplitudes we get (remind that
some numerical factors have already been taken out in (\ref{eq1})):
\begin{eqnarray}
&&\!\!\!\!\!\!\!\!
M_{\lambda_1\lambda_2\lambda_3\lambda_4} = {1\over k_1\cdot k_2 \
k_1\cdot k_3 \ k_1\cdot k_4 \ k_2\cdot k_3 \ k_2\cdot k_4 \ k_3\cdot k_4}
\sum\limits_{cyclic} {\Phi(\lambda_2; 1234) \ \Phi(\lambda_4; 1432)
\over k_2\cdot k_4}\label{eq12}\\
&&\!\!\!\!\!\!\!\!
\times\!\left\{A(\lambda_1; 1234) \varepsilon_{-\lambda_2,\lambda_3,
\lambda_4}^{(1)}(234)
+ B(\lambda_1; 1234) \varepsilon_{-\lambda_2,\lambda_4,\lambda_3}^
{(1)}(243)
+ 2 C(\lambda_1; 1234) \varepsilon_{-\lambda_2,\lambda_3,\lambda_4}^
{(2)}(234)
\right\}\!.
\nonumber\end{eqnarray}
Here the summation extends over simultaneous cyclic permutations of
$(\lambda_1,\lambda_2,\lambda_3,\lambda_4)$ and $(k_1,k_2,k_3,k_4)$:
\begin{eqnarray}
\sum\limits_{cyclic} F(1234) = F(1234) + F(2341) + F(3412) + F(4123).
\nonumber\end{eqnarray}
The $\varepsilon_{\{\lambda\}}^{(1)}$ and
$\varepsilon_{\{\lambda\}}^{(2)}$ amplitudes have been defined in
\cite{b11} (see also \cite{b12}) and we reproduce them in
the appendix.
For the $A,B$ and $C$ functions we have
\begin{eqnarray}
&& A(\lambda_1; 1234) = \varepsilon^{\mu\nu\lambda\sigma} \ k^1_\mu \
q_\nu \
\varepsilon_\lambda^{(\lambda_1)} \  \left[k_3 - {(q-k_1)\cdot k_3\over
(q-k_1)\cdot k_2}\ k_2\right]_\sigma, \nonumber\\[3mm]
&& B(\lambda_1; 1234) = -\varepsilon^{\mu\nu\lambda\sigma} \ k^1_\mu \
q_\nu \
\varepsilon_\lambda^{(\lambda_1)}\  \left[k_4 - {(q-k_1)\cdot k_4\over
(q-k_1)\cdot k_2}\ k_2\right]_\sigma, \label{eq13}\\[3mm]
&& C(\lambda_1; 1234) = i\ \varepsilon^{\mu\nu\lambda\sigma} \ k^1_\mu \
q_\nu \
\varepsilon_\lambda^{(\lambda_1)} \ \varepsilon_{\sigma \mu^\prime \nu^
\prime
\lambda^\prime} \ {(k_1-q)^\mu}^\prime \ {k_2^\nu}^\prime \
{k_3^\lambda}^\prime.
\nonumber\end{eqnarray}
Their explicit expressions are rather cumbersome and are
given in the appendix.

The polarization vectors (\ref{eq8}) and (\ref{eq10})  become ill-defined
for collinear photons. Fortunately, this kinematical region gives a
negligible contribution to the decay width. In fact the corresponding
fictitious kinematical singularities don't cause any considerable trouble
in numerical calculations because the phase factors also vanish for
collinear photons and only the singularities corresponding to the
three simultaneously collinear photons remain.

There are no infrared divergencies in our problem (when the energy of
any photon goes to zero), as is easily seen from the explicit
expressions of the $\varepsilon^{(1)}, \varepsilon^{(2)}$ amplitudes.
The contribution of the muon in the fermion loop can be neglected at
least by a factor $(\omega/m_\mu)^3, \ \omega\sim m/4$, $\omega$ is the
mean photon energy, due to the known low energy behavior of the
light-light scattering amplitude.

The numerical calculations give the result
\begin{eqnarray}
Br (\pi^0\to 4\gamma) \simeq (2.6\pm 0.1) \cdot 10^{-11}.
\label{eq14}\end{eqnarray}
This is about three orders of magnitude below the present experimental
limit.

\appendix
\section{Appendix}
\setcounter{equation}0
\def\theequation{\thesection.\arabic{equation}}

The $\varepsilon_{\{\lambda\}}^{(1)}$ and
$\varepsilon_{\{\lambda\}}^{(2)}$ amplitudes are defined in
our case as
\begin{eqnarray}
&& \varepsilon_{\lambda_2\lambda_3\lambda_4}^{(1)} (234) =
{1\over 2 (q-k_1)^2} E_{\lambda_2\lambda_3\lambda_4}^{(1)} (234)
\nonumber\\
&& \varepsilon_{\lambda_2\lambda_3\lambda_4}^{(2)} (234) =
{1\over 4} E_{\lambda_2\lambda_3\lambda_4}^{(2)} (234)
\label{A1}\end{eqnarray}
and $E^{(1)}$ and $E^{(2)}$ are defined in \cite{b10}. The expressions for
$\varepsilon_{\{\lambda\}}^{(1)}$ and
$\varepsilon_{\{\lambda\}}^{(2)}$  can be found in \cite{b11},
and here we just reproduce them.
\begin{eqnarray}
&&\!\!\!\!\!\!\!\!
\varepsilon_{+++}^{(1)}(234) = {2 (1-\nu_3) (1-\nu_4)\over \nu_3}
+ \left[ {2 (1-\nu_3)^2 (1-\nu_4)
\over \nu_3 (1-\nu_2)} + {2 (1-\nu_3) (1-\nu_4) \over \nu_3^2}\right.
\nonumber  \\
&&\!\!\!\!\!\!\!\!
\left. - {(1-\nu_3)^2\over \nu_3} \right] {\rm ln}(1-\nu_3)
 + \left[{2 (1-\nu_3) (1-\nu_4)\over 1-\nu_2}
 + {(1-\nu_3) (1-\nu_4)\over \nu_4}\right] {\rm ln}(1-\nu_4) \nonumber\\
&&\!\!\!\!\!\!\!\!
 + \left[{(1-\nu_3) (\nu_4-\nu_3)\over 1-\nu_2}
 -{2 (1-\nu_3)^2 (1-\nu_4)\over (1-\nu_2)^2}\right] \
\left({\pi^2\over 6} - {\rm Li}(\nu_3)
 - {\rm Li}(\nu_4)-{\rm ln}(1-\nu_3) {\rm ln}(1-\nu_4)\right),
\nonumber  \\[3mm]
&&\!\!\!\!\!\!\!\!
\varepsilon_{-++}^{(1)} = 0, \ \
\varepsilon_{++-}^{(1)} (234) = -{\nu_2\over \nu_3}
 \varepsilon_{+++}^{(1)} (324) + {\nu_4\over \nu_3}
 \varepsilon_{+++}^{(1)} (342),
\label{A2}\\[3mm]
&&\!\!\!\!\!\!\!\!
\varepsilon_{+-+}^{(1)} (234) = \varepsilon_{+++}^{(1)} (432), \ \
\varepsilon_{-\lambda_2,-\lambda_3,-\lambda_4}^{(1)} (234)
= -\varepsilon_{\lambda_2,\lambda_3,\lambda_4}^{(1)} (234).
\nonumber\\[10mm]
&&\!\!\!\!\!\!\!\!
\varepsilon_{+++}^{(2)}(234) = \left[ {2 (1-\nu_3)\over 1-\nu_2}
 - {1-\nu_3 \over \nu_3}\right] {\rm ln}(1-\nu_3)
 + \left[{2 (1-\nu_4) \over 1-\nu_2}-{1-\nu_4 \over \nu_4} \right]
 {\rm ln}(1-\nu_4)
\nonumber  \\
&&\!\!\!\!\!\!\!\!
- \left[{2(1-\nu_3)(1-\nu_4)\over (1-\nu_2)^2}
+ {\nu_2 \over (1-\nu_2)}\right] \
\left({\pi^2\over 6} - {\rm Li}(\nu_3)
 - {\rm Li}(\nu_4)-{\rm ln}(1-\nu_3) {\rm ln}(1-\nu_4)\right),
\nonumber  \\[3mm]
&&\!\!\!\!\!\!\!\!
\varepsilon_{-++}^{(2)} = -2, \ \
\varepsilon_{++-}^{(2)} (234) = \varepsilon_{+++}^{(2)} (324)
\label{A2a}\\[3mm]
&&\!\!\!\!\!\!\!\!
\varepsilon_{+-+}^{(2)} (234) = \varepsilon_{+++}^{(2)} (432), \ \
\varepsilon_{-\lambda_2,-\lambda_3,-\lambda_4}^{(2)} (234)
= \varepsilon_{\lambda_2,\lambda_3,\lambda_4}^{(2)} (234), \nonumber
\end{eqnarray}
where
\begin{eqnarray}
\nu_2 = 1 - {2\  k_3\cdot k_4\over (q-k_1)^2}, \ \
\nu_3 = 1 - {2\  k_2\cdot k_4\over (q-k_1)^2}, \ \
\nu_4 = 1 - {2\  k_2\cdot k_3\over (q-k_1)^2}, \ \
\nu_2 + \nu_3 + \nu_4 = 2,
\label{A2_1}\end{eqnarray}
\begin{eqnarray}
{\rm Li}(\nu) = - \int\limits_0^\nu {dx\over x} \ {\rm ln}(1-x).
\label{A2_2}\end{eqnarray}
The expressions for the $A,B$ and $C$ functions look like
\begin{eqnarray}
&&\!\!\!\!\!\!\!\!\!\!\!
A(\lambda_1; 1234) = \left[k_1\cdot k_4 + \left(1+{(q-k_1)\cdot k_3\over
(q-k_1)\cdot k_2}\right) k_1\cdot k_2\right] \ [k_1, k_2, k_3, k_4]
   \nonumber\\
&&\!\!\!\!\!\!\!\!\!\!\!
-i \! \left[k_1\cdot k_4 \ q\cdot k_2 \left(k_1\cdot k_3-{(q-k_1)\cdot k_3
\over (q-k_1)\cdot k_2} k_1\cdot k_2\right) +  k_1\cdot k_2 \ q\cdot k_1
\left( k_3\cdot k_4 - {(q-k_1)\cdot k_3\over (q-k_1)\cdot k_2} k_2\cdot
k_4 \right)\right. \nonumber\\
&&\!\!\!\!\!\!\!\!\!\!\!
\left. - k_1\cdot k_2 \ q\cdot k_4 \ \left(k_1\cdot k_3 -
{(q-k_1)\cdot k_3\over (q-k_1)\cdot k_2} k_1\cdot k_2\right) -
k_2\cdot k_3 \ k_1\cdot k_4 \ q\cdot k_1 \right]\ \lambda_1,
\label{A3}\\[5mm]
&&\!\!\!\!\!\!\!\!\!\!\!
B(\lambda_1; 1234) = \left[k_1\cdot k_4 - {(q-k_1)\cdot k_4\over
(q-k_1)\cdot k_2} k_1\cdot k_2\right] \ [k_1, k_2, k_3, k_4] \nonumber\\
&&\!\!\!\!\!\!\!\!\!\!\!
+ i \! \left[k_1\cdot k_4 \ q\cdot k_2 \left(k_1\cdot k_4-{(q-k_1)\cdot
k_4
\over (q-k_1)\cdot k_2} k_1\cdot k_2\right) -  k_1\cdot k_2 \ q\cdot k_4
\left( k_1\cdot k_4 - {(q-k_1)\cdot k_4\over (q-k_1)\cdot k_2} k_1\cdot
k_2
\right)\right. \nonumber\\
&&\!\!\!\!\!\!\!\!\!\!\!
\left. - k_1\cdot k_2 \ k_2\cdot k_4 \ q\cdot k_1 \ {(q-k_1)\cdot k_4
\over (q-k_1)\cdot k_2} -  k_1\cdot k_4 \ k_2\cdot k_4 \ q\cdot k_1
\right]\
\lambda_1, \label{A4}\\[5mm]
&&\!\!\!\!\!\!\!\!\!\!\!
C(\lambda_1; 1234) = \left\{ (k_1\cdot k_2 \ q\cdot k_4 -
k_1\cdot k_4 \ q\cdot k_2) \ [k_1, k_2, k_3, k_4] \
	\lambda_1 \right.\nonumber\\[1mm]
&&\!\!\!\!\!\!\!\!\!\!\!
\left. +i \! \left[(k_1\cdot k_4 \ k_2\cdot k_3
- k_1\cdot k_2 \ k_3\cdot k_4) \ (q^2 \ k_1\cdot k_2 - q\cdot k_1 \
k_1\cdot k_2 - q\cdot k_1 \ q\cdot k_2) \right.\right.\nonumber\\[1mm]
&& - \left.\left. k_1\cdot k_2 \ k_3\cdot k_4 (k_1\cdot k_3 \
q\cdot k_1 + q\cdot k_1 \ q\cdot k_3 - q^2 \ k_1\cdot k_3) \right. \right.
\nonumber\\
&&\!\!\!\!\!\!\!\!\!\!\!
\left.\left. + (k_1\cdot k_2 \ q\cdot k_4 - k_1\cdot k_4 \ k_2 \cdot q) \
(k_1\cdot k_2 \ q\cdot k_3 - k_1\cdot k_3 \ q\cdot k_2)\right]\right\} \
{1\over (q-k_1)^2}.
\label{A5}    \end{eqnarray}
Note that in the squared helicity amplitudes only $[k_1,k_2,k_3,k_4]^2$
appears, and it can be expressed in terms of the scalar products between
photon momenta:
\begin{eqnarray}
[k_1,k_2,k_3,k_4]^2\equiv \left(\varepsilon_{\mu\nu\lambda\sigma}
\ k_1^\mu k_2^\nu k_3^\lambda k_4^\sigma\right)^2
= - \lambda(k_1\cdot k_2 \ k_3\cdot k_4, \ \ k_1\cdot k_3 \ k_2\cdot
k_4, \ \
k_1\cdot k_4 \ k_2\cdot k_3),
\label{A6}\end{eqnarray}
where $\lambda$ is the triangle function.

At last we list the expressions for various scalar products in terms
of the invariant variables used in the phase space integral (\ref{eq2}):
\begin{eqnarray}
\begin{array}{ll}
\displaystyle
k_1\cdot k_2 = {m^2\over 2} (1 + s_2 - s_1 -u_1), &
\displaystyle
	k_2\cdot k_3 = {m^2\over 2} (1 + t_2 - u_1 - u_2),\\[3mm]
\displaystyle
k_1\cdot k_3 = {m^2\over 2} (u_1 - t_2 - s_2),&
\displaystyle
	k_2\cdot k_4 = {m^2\over 2} (s_1 + u_1 + u_2 - s_2 - t_2 -1),
\\[3mm]
\displaystyle
k_1\cdot k_4 = {m^2\over 2} t_2, &
\displaystyle
	k_3\cdot k_4 = {m^2\over 2} s_2,\\[3mm]
\displaystyle
q\cdot k_1 = {m^2\over 2}(1 - s_1), &
\displaystyle
	q\cdot k_2 = {m\over 2}(1 - u_1), \\[3mm]
\displaystyle
q\cdot k_3 = {m^2\over 2}(1 - u_2), &
\displaystyle
	q\cdot k_4 = {m^2\over 2}(- 1 + s_1 + u_1 + u_2).
\label{A7}\end{array}\end{eqnarray}

\end{document}